# Purcell enhanced electroluminescence of a unipolar light emitting quantum device at λ = 10 μm

Marta Mastrangelo[1], Djamal Gacemi[1], Axel Evirgen[2], Salvatore Pes[2], Alexandre Larrue[2], Pascal Filloux[3], Isabelle Sagnes[4], Abdelmounaim Harouri[4], Angela Vasanelli[1], Carlo Sirtori[1]

[1]Laboratoire de Physique de l'École normale supérieure, ENS, Université PSL, CNRS, Sorbonne Université, Université Paris Cité, 75005, Paris, France.
[2]III-V Labs, a joint lab between Thales, Nokia and CEA, 91767 Palaiseau, France
[3]Laboratoire Matériaux et Phénomènes Quantiques, Université Paris Cité, 75013, Paris, France
[4]Centre de Nanosciences et de Nanotechnologies, Université Paris-Saclay, CNRS, Palaiseau, 91120, France.

**Abstract**

Efficient generation of radiation in the mid- and far- infrared relies primarily on lasers and coherent nonlinear optical phenomena driven by lasers. This wavelength range lacks of luminescent devices because the spontaneous emission rate becomes much longer than the nonradiative energy relaxation processes and therefore emitters have to count on stimulated emission produced by linear or non-linear optical gain. However, spontaneous emission is not a fundamental property of the emitter. By engineering metamaterials composed of arrays of nano-emitters into microcavities coupled to patch antennas, we have demonstrated mid-infrared electroluminescent devices emitting a collimated beam with excellent spatial properties and a factor 100 increase in the collected power, compared to standard devices. Our results illustrate that by reshaping the photonic environment around emitting dipoles, as in the Purcell effect, it is possible to enhance the spontaneous emission and conceive efficient optoelectronic light emitting devices that operate close to the thermodynamical equilibrium as LEDs in the visible range.

## 1. Introduction

Light-emitting diodes (LEDs) in the visible range have transformed illumination in all possible aspects: lighting in urban areas, offices and homes, flat-panel displays and they are ubiquitous in dashboards and electronic devices for signalling[1,2]. In most cases, LEDs are semiconductors (either inorganic or organic) in which electrons and holes are injected into the conduction and valence bands, respectively, and recombine through optical transitions across the bandgap, thereby emitting light. The photon emission rate for these devices is determined by the spontaneous emission, $\tau_{sp}$, in a homogeneous three-dimensional space. The rate can be evaluated by using Fermi Golden Rule, which clearly illustrates its dependence on the cube of the emitted photon energy, $\frac{1}{\tau_{sp}} \propto (\hbar\omega)^3$.[3] In the visible, $\tau_{sp}$ is in the order of few picoseconds and therefore spontaneous emission is a very efficient channel for energy relaxation of electrons promoted in the conduction band. In this frequency range, the LEDs have internal quantum efficiency greater than 90% and reach wall plug efficiency in the order of 60-70%.[1] Yet, the spontaneous emission rate scales so rapidly with the photon energy, that in the mid infrared ($\lambda$ = 4 – 20 μm) $\tau_{sp}$ increases of orders of magnitude and the radiative channel becomes very inefficient compared to non-radiative ones. Therefore, LEDs cease to be efficient for wavelengths greater than 1 μm and in the infrared, above 2 μm, only lasers and thermal emitters are used instead.

In this work, we present a unipolar light emitting device operating in the infrared, composed of an array of patch antenna[4,5] nano-emitters that forms an active metamaterial[6], as shown in Figure 1. The semiconductor sandwiched between the two metallic plates is an electrically injected unipolar emitter at 10 μm wavelength (Figure 1).[7] These electroluminescent devices, based on optical intersubband transitions in semiconductor quantum wells, are very inefficient when operate in a mesa configuration,[8] as the rate of non-radiative processes[9] vastly overcomes that of spontaneous emission. However, once embedded into a patch antenna metamaterial, these emitters produce a collimated beam with very low divergence and increase the collected power by two orders of magnitude.

In patch antenna arrays[10] the electromagnetic field is compressed within the microcavity and experiences a sub-wavelength confinement in the perpendicular direction. The cavity mode is resonant with the photon energy, $\hbar\omega_{EL}$, of the electroluminescence from the unipolar emitter and therefore each element of the array acts as a light resonator. Notably, all resonators are connected via a surface plasmon, which gives rise to a super-mode spanning the whole metamaterial. The collective interferential nature of the interaction between all elements of the array enables new frontiers for light manipulation. The shape, size and arrangement of the meta-atoms determine the wavefront of the emitted electromagnetic waves. By controlling the amplitude and phase of the signal at each antenna element, the array can steer the direction of the emitted beam or shape the radiation pattern. The design and operation of these metamaterial devices are therefore based on the principles of wave constructive/destructive interference and array theory.

The microcavity arrays have been already exploited for realising very efficient metamaterial detectors, QWIPs[4] and QCDs[11], and modulators[12]. Moreover, their hybrid nature, that combine antenna and microcavity effects, has been also used for fundamental studies, in particular to enhance light-matter coupling and explore the ultra-strong coupling regime[13,14]. In the THz domain, patch antenna arrays have been used to realise phase shaping[15], perfect absorbers[16] and single-mode lasers with low beam divergence[17].

## 2. Sample description

A scanning electron microscope image of the device is reported in Figure 1: it consists of an array of microcavities with lateral size *s* and thickness *H*, placed in an array with periodicity *p*. The cavity size *s*, whose value is close to 1.4 μm, is set by $2ns = \lambda$, where *n* is the modal refractive index. The other parameters are set as $H = 0.75$ μm and $p = 7$ μm. Each array has a total size of approximately 100 – 150 μm in size. The active region of a quantum cascade emitter is embedded between the metal layers. It consists of an InGaAs/AlInAs heterostructure designed to operate at ∼ 9 μm, whose bandstructure is also illustrated in Figure 1. The top and bottom metal layers act also as electrical contacts for current injection.

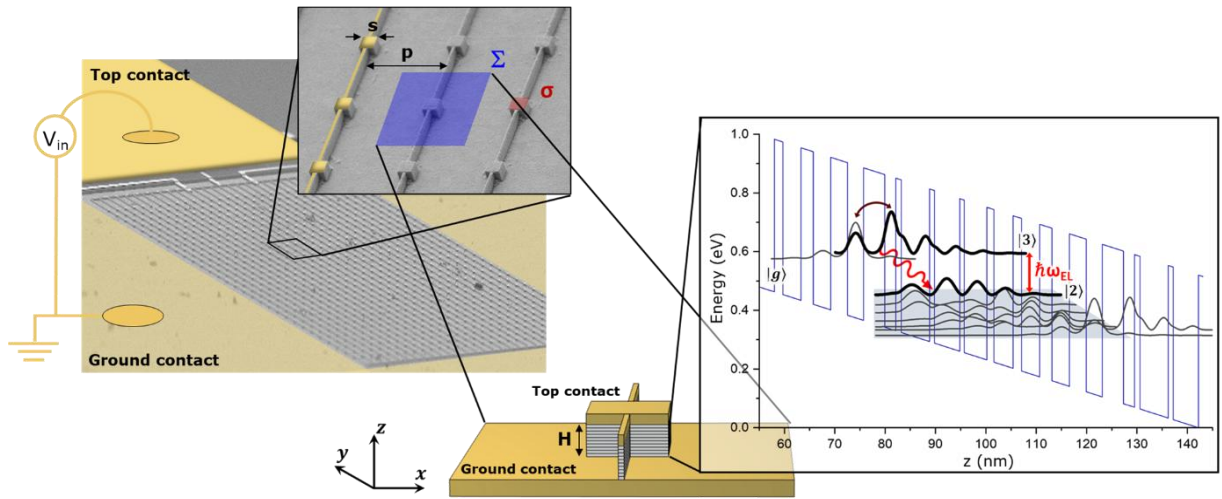

*Figure 1 Scanning Electron Microscope (SEM) image of the patch-antenna array. The QCL active region is embedded within double-metal microcavities. The top and ground metal layers, highlighted for clarity, provide electrical connections to the device. The inset offers a magnified view of the cavities, illustrating the distinction between the electrical area σ = s² and the optical collection area Σ = p², where s is the size of the patch cavity and p the periodicity of the array respectively. The electrical area is limited to the cavity dimensions, while the optical area extends beyond, encompassing the inter-cavity spacing. The panel on the right presents the conduction band profile and relevant square moduli of the electronic wavefunctions for one period of the quantum cascade emitter inserted into the microcavities. The electroluminescent transition occurs between levels |3> and |2> at an energy $\hbar\omega_{EL}$ ~ 130 meV (10 μm wavelength). The transition has a strong Stark shift due to its diagonal nature.*

The fabrication flow of the device begins with wafer bonding and substrate removal to create the buried metal layer that serves as the bottom contact of the patches. A Ti/Au bilayer is deposited on the InGaAs/AlInAs heterostructure, which is then bonded by thermocompression to a Ti/Au-coated host substrate. The original InP substrate is subsequently removed through selective etching in hydrochloric acid, leaving the active region transferred onto the metal-bonded host substrate. An insulating layer of silicon nitride ($Si_3N_4$), 300 nm thick, is then deposited to electrically isolate the top contact. This layer is patterned and etched to define dielectric pads where the top electrode will later be placed. The patch antennas are defined by electron-beam lithography, followed by metal deposition (Ti/Au/Pt) and lift-off. This procedure simultaneously forms the top metal contact (Ti/Au) and the platinum (Pt) metal mask used during dry etching (ICP-RIE with $CH_4/H_2$) to define the antenna profiles.

The patch antenna resonator array was first characterized by performing reflectivity measurements by using a Fourier Transform infrared spectrometer (FTIR) with a Cassegrain microscope objective (Bruker Hyperion II), which allows to focus infrared light onto an individual device. The measured quality factor of the fundamental cavity mode is $Q_{cav}$ = 14.

Under an appropriate bias, electrons populate the excited state of the radiative transition and emit photons by spontaneous emission. The emitted photons arise from the intersubband dipole that oscillates perpendicular to the plane of the QWs. In order to populate the cavity mode, a π phase shift is impressed on them from one edge to the opposite of the cavity. The electromagnetic field of the mode couples strongly with the dipole of the antenna which releases photons in the free space. Notably, the patch antenna cavity allows the microscopic intersubband dipoles, aligned in the z-direction, to do surface emission, rather than emitting in the plane. Furthermore, the array improves emission directionality, as the antennas interact constructively only along one direction and destructively in all the others, providing very low beam divergence.

The emission properties of our device arise therefore from the complex interplay between the properties of the emitter and those of the photonic structure. Figure 2.a shows electroluminescence spectra, for different injected currents from a device processed into a mesa structure, thus without cavity effect. The energy peak of the electroluminescence, $E_{EL}$, shifts for different injected currents due to the inherent Stark shift of this active region design. The energies $E_{EL}$ are plotted, in Figure 2.b, against the reflectivity spectra of cavities with different sizes $s$ (from 1.3 to 1.5 µm in steps of 50 nm). From Fig. 2.b, it is clear that by modifying the injected current (applied bias) it is possible to drive the peak of the electroluminescence, $E_{EL}$, in and out the resonance of the cavity mode.

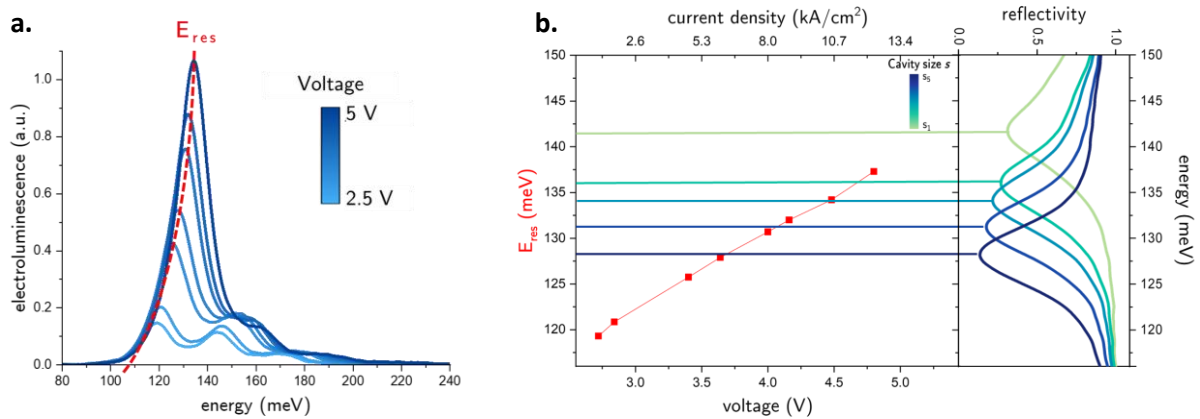

Figure 2 a. Electroluminescence spectra measured on a device in a mesa geometry (i.e. without cavity) at different applied voltages at 78 K. b. Energy resonances $E_{res}$ extracted from the electroluminescence spectra without cavity (left panel) as a function of the applied voltage. The right panel and reported in b. are retrieved and reported together with the position of the patch cavity resonances represented by the horizontal lines, deduced from reflectivity measurements of different lateral patch sides s, spanning from 1.3 µm to 1.5 µm.

### 3. Results on patch antenna metamaterial

The array configuration allows shaping the emission beam in real space, reducing its angular divergence in the far-field. Indeed, the directivity of the emission is directly proportional to the array size, given by the product of the number of emitters in the array and their inter-element spacing[18]. Our first characterisation was therefore the measure of the far-field profile at different distances to estimate the beam divergence of our devices. Figure 3 illustrates a sketch of the setup used for far-field reconstruction (a) and a typical profile with the relative beam divergences (b). The measurement is performed at 78 K by applying a bias of 4.2 V on the device in pulsed mode ($t_{pulse}$ = 670 ns). The emitted light is directly collected by a mercury cadmium telluride (MCT) detector placed in front of the emitter with no converging optics in between. The detector is mounted on a translational stage used to precisely move in the xy plane, parallel to the metasurface, to reconstruct a 2D intensity map of the emission from which the beam divergence $\theta_{div}$ can be estimated. To ensure consistency, the measurements are repeated at various distances along the z-axis from the emitting sample, always remaining in the far-field zone. The far-field intensity distribution measured for a 20 × 20 array (140 x 140 µm²) at 78 K is reported in (b). Cross-sectional profiles along the x and y directions through the beam centre reveal a spatial broadening (taken as their FWHM) of *Δx* = 0.59 mm and *Δy* = 0.68 mm, respectively. Given the distance between the emitter and the MCT detector, z = 50 mm, the spatial broadening can be converted to angular divergence using the relation:

$$\theta_{div,x(y)} = 2\,arctan\left(\frac{1}{2}\frac{\Delta_{x(y)}}{z}\right)$$

obtaining $\theta_{div,x}$= 0.67° and $\theta_{div,y}$= 0.77°. The remarkably low beam divergence is a direct result of the array configuration, which sets the angular divergence inferior to 1°. Therefore, the device produces a naturally collimated beam without the need for external collimation optics. This self-collimation is a significant advantage over edge-emitting devices, which typically require additional optical components to produce a similar beam profile. Such a narrow beam is unprecedented. The record value established for THz QCLs based on patch-antenna metamaterial is 2°.[6] Such a directional emission clearly confirms that the individual dipoles associated with each microcavity are phase-locked by the super-mode, spanning the whole metamaterial, induced by the surface plasmon on the bottom metallic layer.

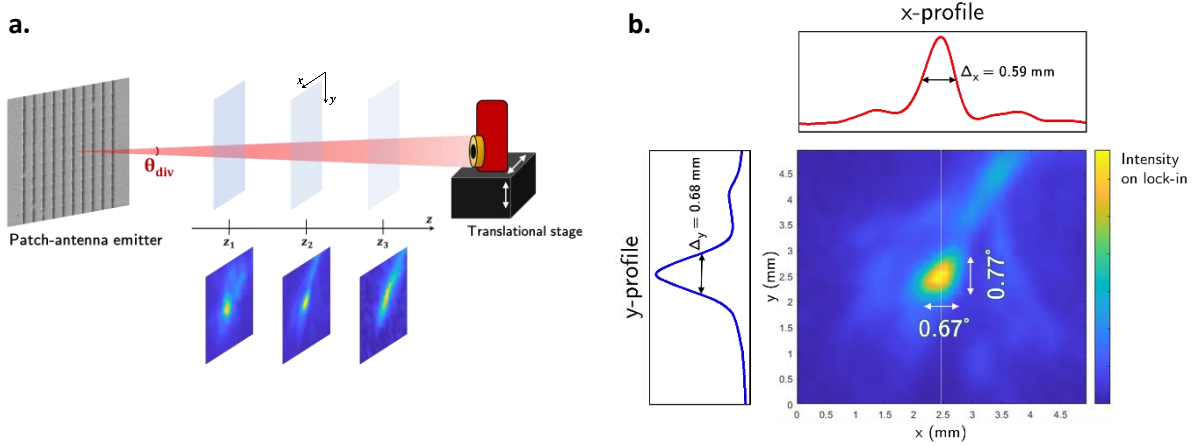

*Figure 3 a. Experimental setup for characterising far-field radiation from patch antenna arrays. The MCT detector, placed in the far-field zone of the device with no converging optics in between, directly collects the emitted signal. To map the emission profile in the xy plane parallel to the metasurface, the MCT is mounted on a motorised translational stage and the measurement is repeated at different distances z from the emitter to confirm the consistency of the obtained results. This enables to reconstruct the spatial far-field profile and quantify the relative beam divergence, as shown in (b), where the spatial intensity distribution with corresponding x and y profiles of the far-field emission of a 20×20 patch-antenna array (s = 1.4 µm) is reported. The broadening, respectively of 0.59 mm and 0.68 mm, are used to compute the angular divergence in the two directions, obtaining $\vartheta_{div,x}$ = 0.67° and $\vartheta_{div,y}$ = 0.77°. The measurement is performed at 78 K applying a bias 4.2 V on the device in pulsed mode ($t_{pulse}$ = 670 ns), placing it at a distance of 50 mm from the MCT detector.*

To complement the characterization of the devices embedded in the metamaterial, we measured the electroluminescence spectra, as well as the calibrated emitted power and the quantum efficiency. In Fig. 4.a we show two electroluminescence spectra at 4.5 V, one from a mesa device and the second from a 10 × 10 patch-antenna array with lateral patch size s = 1.4 µm. We can observe that the patch cavity effectively "filters" the emission, resulting in an evident spectral narrowing of the device in cavity, whose linewidth, $\Delta E_{EL}$, reduces of factor 2 compared to the mesa device. Note that the quality factor of the electroluminescence spectrum $Q_{EL} = 18$ is larger than that of the bare cavity extracted from reflectivity spectra ($Q_{cav} = 14$). This is an indication that the patch cavity does not only act as a spectral filter, but also as a resonator for the modal gain.

The emitted power, $P_{opt}$, was carefully measured using a calibrated photodetector. Specifically, Fig. 3.b reports the emitted power as a function of the current injected in the two devices. The optical power rises much faster in the patch-cavity device and at 80 mA reaches 2.9 µW, whereas in the mesa is 0.03 µW, thus an enhancement of about two orders of magnitude. The slope of the curves is related to the emission quantum efficiency $\eta_{QE}$, which is the ratio between measured emitted photons and injected electrons:

$$\eta_{QE} = \frac{N_{photons}}{N_{electrons}} = \frac{P_{opt}/\hbar\omega_{EL}}{I/q}$$

In our device, we have measured a value of 2.25×10⁻⁴, which is much higher compared to typical values obtained in light sources relying on spontaneous emission.[8,19]

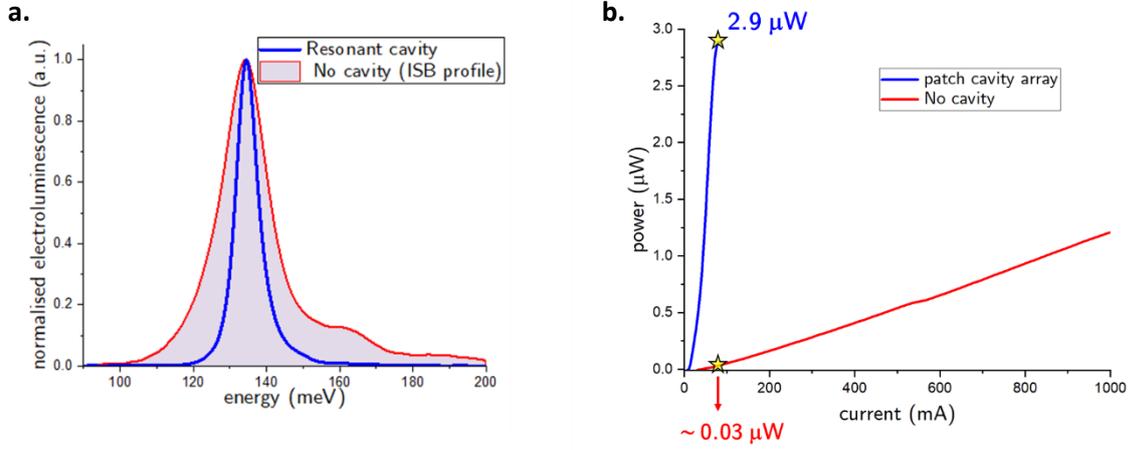

*Figure 4 Emission properties of a 10×10 patch-antenna array with lateral patch size s = 1.4 µm compared with a non-cavity device at 78 K, specifically the spectral features (a) and the emitted power as a function of the injected current (b) are shown. The electroluminescence spectra in (a) are measured on both cavity (blue line) and non-cavity (grey-filled, red line) devices at the same operating voltage of 4.5 V, corresponding to the point where the ISB transition is aligned with the cavity resonance. Panel (b) compares the intensity of the emission, reporting the measured power as a function of injected current. The slopes of the curves correspond to the emission quantum efficiency η_QE, which exhibits an enhancement of 2 orders of magnitude in the patch cavity case.*

Such an enhancement in the emission featured in patch-antenna devices can be ascribed to the combination of a greater collection efficiency and the Purcell effect. Indeed, in the patch antenna device light is surface emitted in a collimated beam. Moreover, the ensemble of the dipoles can couple to the only mode of the cavity which imposes a sizeable modification of the spontaneous emission rate, called the Purcell effect[20,21]. This phenomenon can strongly increase the rate when the matter and cavity oscillators are resonant or, on the contrary depressed, if not extinguished, when the optical mode is far from the resonance of the dipoles. To quantify the enhancement of the emission rate in a cavity with volume $V_{cav}$ we use the Purcell factor, $\mathcal{F}_P$, which is defined as the ratio between spontaneous emission rate in the cavity $\Gamma_{cav} = 1/\tau_{sp}^{cav}$ with respect to that in a homogeneous medium $\Gamma_0 = 1/\tau_{sp}$. This factor can be expressed as a function of the spectral detuning between the emitted photon energy and that of the cavity mode, $\hbar\omega_{EL} - E_{cav}$,:[22]

$$\mathcal{F}_P = \frac{\Gamma_{cav}}{\Gamma_0} = F_P(Q, V_{cav})\mathcal{L}(\hbar\omega_{EL} - E_{cav}, \Delta E)$$

with $F_P(Q, V_{cav}) = \frac{3Q(\lambda_{cav}/n)^3}{4\pi^2 V_{cav}}$ and $\mathcal{L}(\hbar\omega_{EL} - E_{cav}, \Delta E) = \frac{\Delta E^2}{4(\hbar\omega_{EL} - E_{cav})^2 + \Delta E^2}$

$F_P$ is the Purcell coefficient, the highest rate of spontaneous emission that can be delivered, while $\mathcal{L}$ is the Lorentzian that reduces it due to the detuning between cavity mode and peak emission.

Here $Q$ is the quality factor computed as the harmonic sum of the quality factors of the electroluminescence, $Q_{EL} = \hbar\omega_{EL}/\Delta E_{EL} = 9$, and the cavity $Q_{cav} = 14.4$, $\lambda_{cav} = \frac{hc}{E_{cav}}$ is the resonant wavelength of the microcavity and $\Delta E = \Delta E_{EL} + \Delta E_{cav}$ is the linewidth of the Lorentzian.[23] Using all these values, we calculated $F_P$ = 7. Notably, due to the Stark effect of our emitter, the detuning,

$\hbar\omega_{EL} - E_{cav}$, varies and therefore $\mathcal{F}_P$ becomes a function of the applied bias, thus of the injected current, $\mathcal{F}_P(J)$. Note that in our system all dipoles oscillate in the same direction of the electric field in the cavity (Fig. 1c) and therefore no angle correction is needed in the expression of $\mathcal{F}_P$.

We model the emission in the cavity by using adapted quantum cascade laser rate equations [24] that include the Purcell factor on all the radiative transitions, associated with both spontaneous and stimulated emission:

$$\frac{dn_3}{dt} = \frac{J}{q}\frac{1}{L_p} - \frac{n_3}{\tau_3} - \mathcal{F}_P\left(\frac{n_3}{\tau_{sp}} + \frac{\Delta n}{\tau_{st}}\right)$$
$$\frac{dn_2}{dt} = \frac{n_3}{\tau_{32}} - \frac{n_2}{\tau_2} + \mathcal{F}_P\left(\frac{n_3}{\tau_{sp}} + \frac{\Delta n}{\tau_{st}}\right) \quad (1)$$
$$\frac{dS}{dt} = -S\Gamma_{tot} + \mathcal{F}_P\left(\frac{n_3}{\tau_{sp}} + \frac{\Delta n}{\tau_{st}}\right)$$

Here $n_i$ is electronic volume density of the $i^{th}$ subband; $J$ is the injected current density; $q$ is the elementary charge; $L_p$ is the length of a period of the cascade; $\tau_i$ is the non-radiative lifetime of an electron on the subband $i$; $\tau_{st}$ is the stimulated emission time, $\Delta n = n_3 - n_2$ is the population inversion; $1/\tau_{32}$ is the non-radiative rate associated with the transition from subband 3 to subband 2; $S$ is the photonic volume density; $\Gamma_{tot}$ is the total losses, or equivalently the inverse of the lifetime of the photon in cavity, given by the sum of the radiative, $\gamma_R$, and non-radiative, $\gamma_{NR}$, contributions.

These rate equations have been established considering that the spontaneous emission coupling factor is equal to one, as in patch antennas emitted photons couple with a single optical mode. In other words, all the emitted photons, weather they are issued from spontaneous or stimulated emission, are generated in the same optical mode.

The third equation, solved at steady state, sets a condition for the emission in the microcavity: the radiative emission rate (including both spontaneous and stimulated emission) should equal the cavity losses. From this condition, we can write the photon density as a function of the upper state population and of the population inversion:

$$S = \mathcal{F}_P \frac{n_3}{\tau_{sp}} \frac{1}{\Gamma_{tot} - \Delta n\, \sigma_V \mathcal{F}_P}$$

Here $\sigma_V$, is the modal gain rate ($\sigma_V S = 1/\tau_{st}$). In the sub-threshold operation regime, the stimulated and the spontaneous emission time are much longer than the non-radiative one. Consequently, the upper state population $n_3$ and the population inversion can be calculated in the cold cavity approximation. The emitted photon density is then written analytically as a function of the current density as:

$$S = \frac{\mathcal{F}_P}{\tau_{sp}} \frac{J}{qL_p} \frac{\tau_3}{\Gamma_{tot} - \sigma_V \mathcal{F}_P \frac{J}{qL_p}\tau_{eff}} = \mathcal{F}_P \frac{\tau_3}{\tau_{sp}} \frac{1}{\tau_{eff}\sigma_V} \frac{J}{\frac{\Gamma_{tot}}{\sigma_V}\frac{qL_p}{\tau_{eff}} - \mathcal{F}_P J}$$

with $\tau_{eff} = \tau_3\left(1 - \frac{\tau_2}{\tau_{32}}\right)$. In order to highlight the influence of the Purcell factor on the threshold condition, we can introduce the threshold current density of a quantum cascade laser without microcavity ($\mathcal{F}_P=1$) and neglecting the effect of spontaneous emission on $n_3$:

$$J_{th} = \frac{\Gamma_{tot}}{\sigma_V}\frac{qL_p}{\tau_{eff}}$$

The emitted photon density then becomes:

$$S = \frac{\tau_3}{\tau_{sp}} \frac{1}{\tau_{eff}\sigma_V} \frac{J}{J'_{th} - J} \quad (2)$$

where the microcavity laser threshold is $J'_{th} = J_{th}/\mathcal{F}_P(J)$. The formula illustrates the strong dependence of the photonic density, $S$, on the Purcell factor that varies the value of the effective threshold current density, $J'_{th}$, as a function of the detuning of between the cavity and the emitter. This model can be used also to estimate the emitted power, $P_{opt} = S\,\hbar\omega_{phot}\gamma_R V_{cav}$, and thus to fit the light output versus current density curves obtained experimentally. Figure 5 illustrates the experimental curves (blue line) for patch arrays with different cavity sizes, *s*. The data are very well reproduced by our model using eq. 2 (red line), where the only fitting parameter left is $J_{th}$. The results for all different device sizes are very consistent and estimated the threshold for laser action at ∼ 25 kA/cm². This very high threshold current density is in agreement with the optical losses of the cavity which are 10 times higher than that of a ridge laser and set the photon lifetime $\tau_{loss} = \frac{1}{\Gamma_{tot}}$ to 1 ps. Note that the about 80% of these losses are due to the radiative rate which is imposed by the patch antenna that couples the light out of the cavity very efficiently.

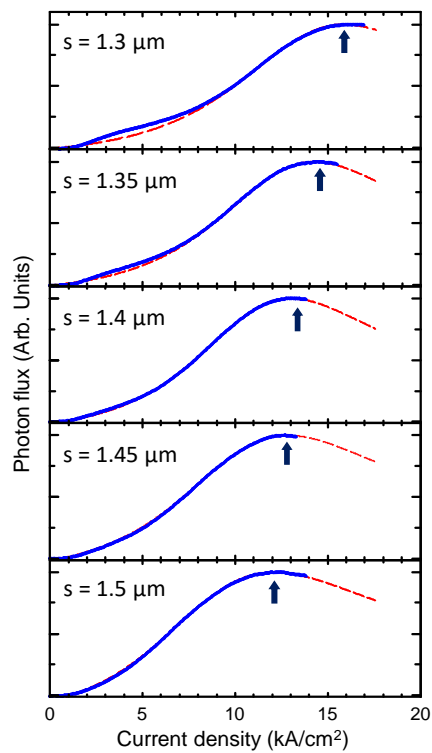

*Figure 5 Normalised photon flux as a function of the current density J for samples with different lateral patch size s. The graphs compare the experimentally measured curves (in blue), with the theoretical formula derived by solving QCL rate equations in sub-threshold regime (red dashed line) (equation 2). The black arrows indicate the position of the photon flux peak that shifts with the patch size s. The threshold current $J_{th}$, is a fitting parameter and is estimated to be 25 kA/cm² for all sample sizes.*

## 4. Conclusions

We have demonstrated a new type of unipolar quantum optoelectronic device based on an array of microcavities where light emission forms a collimated beam with an intensity enhancement due to the Purcell effect. Indeed, the characterisation of the far-field radiation pattern shows surface emission with a very low divergence output that optimises the collection efficiency of the emitted radiation. These properties reduce the need of optical elements and therefore are a great advantage to design and realise optical systems. Moreover, the microcavity improves the spectral purity of the emission and, by adjusting its size, the wavelength can be easily tuned.

Our results are supported by a model that describes the Purcell enhancement including both spontaneous and stimulated emission mechanisms and indicate that the threshold for achieving laser action in these light sources is totally limited by radiative losses.

Given the thorough optimisation conducted on the metamaterial geometry, further reduction of losses within the current configuration appears unlikely. However, our preliminary investigations have unveiled a promising alternative strategy: exploiting higher-order cavity modes beyond the fundamental one. Electromagnetic simulations have demonstrated that an adequate mode engineering can yield cavity modes with higher quality factors. These higher-Q modes inherently exhibit reduced losses and, consequently, lower laser thresholds. This approach can potentially overcome the limitations imposed by the fundamental mode losses without necessitating fundamental changes to the existing metamaterial structure and maintaining the benefits of our optimised metamaterial design.

Acknowledgements

This work was supported by ENS-Thales Chair, PEPR Electronique and the French RENATECH network. The authors acknowledge fruitful discussions with Jérôme Faist and Jaime Gomez Rivas.